\documentclass[11pt]{article}
\usepackage{a4wide}

\usepackage{amsmath, amssymb, amsthm, amsfonts, dsfont, bbm}
\usepackage{enumerate}

\def\phi{\varphi}

\newcounter{qcounter}                                           

\begin{document}

\title{Blind proxy voting}
\author{Zuzana Hanikov\'a\\
Institute of Computer Science\\
 Czech Academy of Sciences\\
{\tt hanikova@cs.cas.cz}}
\maketitle

\begin{abstract}
A secret ballot mechanism that enables voting in absence is proposed. 
It amends standard vote collection methods that use ballot box as anonymizer, 
adding the option for absent voters to vote by a proxy blinded to the content
of the ballot paper. 
Votes are cast under unique and hidden identification numbers
generated solely for the purpose of that election.
Voters prepare their ballot papers from scratch and submit them to 
the tallying authority in two parts via separate routes,
 each part being meaningless without the other.
 
\end{abstract}

\section{Introduction}

This document offers a simple vote collection method that 
enables absentee voting while meeting a secret ballot requirement.
The method is a modification of a standard collection method
where voters arrive at polling place to cast their votes by marking their ballot papers appropriately, 
 deploying a ballot box
and an electoral roll.
It adds the possibility of voting by proxy but blinds the proxy to the 
information he is transmitting for his principal.\footnote{The term \emph{proxy voting}
can convey varied nuances of meaning; we use it here for the act of delegating the right to vote
to another voter (or more generally, another person) with a clear indication of voting intent.}
The system was designed with a closed community in mind, such as an organization. Nevertheless,
it extends, on a few assumptions discussed below, to any large-scale elections where the necessity
arises to allow participation for absent voters on fair and comparable terms.

Without trying to define or explain the somewhat vague notion of an ``electronic election'',
we remark upfront that this document does not propose one.
All agents in the proposed system are intended to be persons;
it does not rely on sophisticated mechanical or electronic devices or software, 
and can be implemented with a minimalist traditional equipment (pen, paper, and a ballot box), 
although it is expected that voters and authorities
do have access to (non-encrypted, non-anonymous) usual communication means, such as post,
email, or phone. 
Although electronic voting systems are likely to enable absentee voting,
we try to contribute, as a side issue, to the perception that the former is not necessitated by the latter.

The proposed system is homogeneous in that it does not impose a  difference  
between those voters who are able, and those who are unable, or choose not, to come to polling place. 
Each voter can decide for herself whether she will attend in person or vote by proxy, 
without announcing her decision to any authority; 
only the voter and her appointed proxy, if there is one, know what she has opted for.\footnote{If a voter
is known to be away from the polling place (travelling, for example), then of course
those in the know are aware that voting by proxy is her only option. 
However, numerous voters may opt for using a proxy who might well be able
to come to the polling place, and for any of these voters, the system leaves no trace of whether or not they used a proxy.} 
The design expects all voters to generate their ballot papers, and all voters also
use analogous routes to deliver them for tallying, 
regardless of whether they vote in person or by proxy.
This is highly desirable since any difference in the nature of ballot papers or
in the manner of vote casting might implicate a (possibly narrow) group of voters.

Prior to voting, voters receive unique identification numbers (IDs) from a registering authority.
A voter creates her own ballot paper; there is no need for delivery of pre-printed ballot forms.
Each  voter distributes the content of her ballot paper into two parts, 
each of them meaningless without the other,
both bearing that voter's ID.
The two parts are submitted via routes that only meet at the tallying point; 
one part is submitted via the registering authority, uniformly across voters;
the other part is submitted in person or by  proxy on election day.
The full content of the ballot paper is divulged to neither route,
and the information about who cast which vote is not available to any single party;
 both can be restored by misdemeanour of more than one of the parties.

The  ballot paper is distributed using a ``grid--grille'' method (\cite {cardangrille}).
This ancient system of private communication is listed as a cryptographic method, 
but it is intuitive in the sense that a voter (generally, parties who wish to communicate) can use it 
without extra education.

\subsection{Task}
\label{ss:task}

An unambiguously defined pool of voters (persons with a right to vote) is assumed as a given.
For our exposition, we assume $m$ candidates run to fill $n$ seats,
and each voter can vote for an arbitrary subset of the set of 
candidates\footnote{We note in Section \ref{section:modif} that the design can be adapted to a number of electoral systems.}
(in particular, it is possible to vote for more candidates than there are seats).
The candidates are some of the voters. Prior to election day, 
a list of candidate names is fixed and made known.
The result of an election is an an assignment of a natural number to each candidate name, 
specifying how many for-votes that candidate obtained in the election, and the implied order
of candidates.

Further, it is assumed that election day and polling place 
have been specified. An electoral committee,
that may include some of the voters (but presumably not the candidates) has been appointed
with the task of collecting and counting votes.

For simplicity, it is also assumed that the election takes place in a particular district,
where there is only one instance of each of the authorities controling the election and props employed:
 the committee, the electoral roll, the polling place, the ballot box, etc. 
This corresponds to the intended result of the election, as detailed above.
Multiple instances of this setup can naturally operate independently of each other.

\subsection{Standard method}

We briefly outline the ``standard vote collection method'' that we seek to modify.
The method expects voters to come to the polling place in order to be 
able to cast a vote.\footnote{Optionally, voters unable to come are able to vote by proxy, on pain of revealing their intent. In other such systems,
ballot box can be brought to voters who are unable to attend at polling place.
In order to do that, the authorities need to know their location, and that needs to be easy to reach.}
The method is based on stringent control of ballot box access:
to guarantee that only registered voters cast a vote, 
and that each voter only votes once, 
each voter's identity is checked against an electoral roll 
(sometimes with the obligation of signing their name) upon casting a vote,
thereby waiving further access to the ballot box. 

The strength of the above mechanism is its simplicity:
a voter need not exert any effort beyond arriving and marking her ballot paper appropriately; 
the marks are usually foolproof.
Tallying is straightforward. 
The ballot box, albeit somewhat crude, is considered a credible anonymizing device, 
breaking the link between the voter and her marked ballot paper.

\section{Proposed system}

The proposal only concerns vote collection; 
the superposed parts of an electoral system are kept intact and not discussed here.
No extra equipment is necessary other than that already in use in the standard system
(except for communication means).

A registering authority (RA) is considered throughout as one of the roles in the system. 
It is necessary that the identity of all persons involved in the registering is publicly known,
so that they are not considered as proxies. 
It is furthermore assumed that the role of the registration authority and that of the electoral committee (EC) are disjoint.
Lastly, RA ought to be collectively easy to reach. 

Prior to voting, each voter receives from the RA in private a unique identification number (ID),
which is a large natural number chosen at random.
The assignment of ID's is kept hidden.
Each vote is then cast under that voter's ID.

The information contained in a marked ballot paper is divided into two parts, 
here called the \emph{grid} and the \emph{grille}; 
we reserve \emph{ballot paper} for the pair.
One can think of the grid and the grille as two columns on a sheet of paper. 
The grid column is a list of all candidate names and possibly other names or strings, 
their number and order deliberately and privately chosen by the voter. 
The grille column is a $\{0,1\}$-vector of the same length as the list, 
specifying which candidates are voted for from that ballot paper: $1$ means a for-vote.
The grid and the grille are delivered to the electoral committee separately, 
both bearing the respective voter's ID. 
Upon tallying the results, the committee learn the valid IDs used in the election;
using these, each grid is paired with its corresponding grille, yielding the complete ballot papers.

Voters who cannot or will not come to polling place can appoint proxies to communicate their vote, 
in a manner described further on.

In the standard method, it is trivial that only voters can access the ballot box and insert ballots.
In the proposed modification, it is necessary to define in advance not only who the voters are, but also
who can act as a proxy. Typically proxies are found among the voters, but aberrations from this expectation are possible.
Whatever the final arrangement, all voters and all persons who might serve as proxies 
need to be allowed to insert papers in the ballot box on the day of the election.

Ballot box access is not as strictly controlled as in the standard mechanism.
In particular, voters (and proxies, if they extend the set of voters)  can repeatedly insert 
sheets of paper in the ballot box, and the contents of any paper must not be questioned or inspected until after election day;
some of these papers are later evaluated as ballot papers, others may be considered noise and discarded. 
On election day, the committee only make sure that \emph{nothing is taken out} from the box.
Thus, instead of filtering voters for ballot box access, the committee filters ballot papers by the validity of their IDs.


\subsection{Blind proxy voting in steps}

\noindent
{\bf Roles:} 
\begin{itemize}
\item candidates; 
\item voters; 
\item electoral committee (EC); 
\item registering authority (RA); 
\item proxies.
\end{itemize}

\smallskip\noindent
{\bf Prerequisites:} a list of candidates, an electoral roll, 
a list of eligible proxies\footnote{when different from the electoral roll; for example, if persons without a right to vote are able to act as proxies};
fixed election day and polling place, ballot box.

\smallskip\noindent
The following sequence of steps replaces the standard vote collection and reading.

\smallskip\noindent
{\it  Step 1: registration of voters.}
The RA assign to each voter a unique ID.
Her ID can be delivered to a voter in person, by post or by email.
The assignment of IDs to voters is kept hidden throughout, including anytime after the election.
Moreover, the information on which IDs are being used in an election 
is not made public until after all votes have been collected (step 3).

\smallskip\noindent
{\it  Step 2: preparing ballot papers.}

- Victoria the voter chooses a set of names that consists of all candidate names and 
optionally also other names or strings.\footnote{Twice the number of candidates should be sufficient;
cf. section \ref{sssection:qa}. Should Victoria, on the other hand, omit some of the real candidate names,
but otherwise deliver a valid ballot, there is an option of interpreting this as her not voting for the omitted
candidates.}

- Victoria lists the set of names in order of her choice. This is the grid part of Victoria's ballot paper. 

- Victoria prepares the grille part of her ballot paper, which consists of a characteristic vector of the voted-for 
names in her grid: 1 is interpreted as a vote-for and 0 as a not-vote-for the name on that line. 
This column spans also the names that are not candidate names (the values will be ignored in the tally). 
She records her ID onto the grille part.

- Victoria counts the number of $1$'s in her grille (the checksum).
She then submits back to RA the grid and the checksum under her ID, 
again in person, by post or email, keeping the grille part.\footnote{Presumably
the RA will set a deadline for voters to submit their grids, possibly with some publicly distributed reminders.}

- Victoria makes up her mind as to visiting the polling place or voting by proxy. 
If she opts for the latter, she finds a peer willing to act as her proxy and 
 able to confirm he will visit the polling place.\footnote{In order to determine this, 
  the election day needs to be fixed by the time this step takes place. 
Victoria is not expected to opt for a member of RA or EC as a proxy, since either would combine
her identity with her ballot paper.}
She then emails or otherwise communicates her grille with her ID number to her proxy.
The proxy records Victoria's grille and ID to a sheet of paper in order to be able to insert it in the ballot box.

\smallskip\noindent
{\it Step 3: collection of ballot papers.}
Election day starts with the EC sealing the empty ballot box. 
During that day, voters and proxies can insert the grille parts of ballot papers. 
The committee attend the ballot box throughout that period,
 supervising that nothing is taken out.
Each voter or proxy can insert one or more sheets of paper.  
The committee are not expected to inspect the number, nature, or content of the inserted papers.

Only after the vote casting has ended, 
RA deliver to the EC the list of valid IDs and 
all the grid parts of ballot papers with the checksums, each submitted under its valid ID. 
(In particular, the committee learn the valid IDs at that point,
but \emph{not} the assignment of IDs to voters.)

\smallskip\noindent
{\it Step 4: reading ballot papers and announcing results.}
The  committee open the ballot box. 
All sheets of paper without a valid ID are discarded.
If two or more grilles bear the same ID, that ID is invalidated and any papers with that ID are disregarded henceforward.
If either the grid or the grille part is missing for a valid ID, that ID is invalidated as well.
If for a delivered grille  with a given ID, the sum of 1's differs from the checksum for that ID (submitted with the grid),
or otherwise the grid and the grille for a given ID do not match,
that ID is invalidated.
Subsequently, for all IDs that are still valid, 
the committee assemble the grid-grille pairs;
malformed ballot papers are discarded.
Finally, the committee obtain a set of complete and valid ballot papers.

For each ballot paper with a valid ID, all names other than candidate names are ignored. 
The committee count the total number of for-votes for each candidate. 
This, and the total number of valid delivered ballot papers, constitutes the election result.

With the result, the committee publish 
 the IDs that have beem generated by the RA, out of these, the IDs that have 
taken part in the election, and out of these, the invalidated IDs, if any, 
each with its respective status.

The assignment of ID's to voters (not the ID's in use), as established by the RA prior to the election,  can be deleted permanently after the election is over.

\medskip
Let us look at  the timespan and concurrency options of the subsequent steps.
First of all, the registration of voters by the RA (step 1)  is independent of  
 other preparatory election stages, such as the registration of candidates, campaigning and 
 the presentation of manifestos etc., and can therefore happen in parallel. 
The duration is proportional to the size of the electoral roll,
but in reasonably sized organizations or districts, it might take one or two weeks.
 
The preparation of ballot papers may also need up to one week collectively, although individually, it may take only a few minutes.

The voting (step 3) can span a day---since it can be assumed 
that the possibility of voting by proxy gives everyone, including voters who cannot come to polling place,
an equal chance of participating.

Finally, the reading of ballots is again proportional to the size of the roll, and inversely to the size of the committee. 
Under a reasonable ratio of number of voters to committee headcount, 
 and no major attempt at stalling the election, it ought not to take significantly more than 
another day to count and report the results.

\subsection{Discussion}

As already advertised,  both the standard  method
and the proposed modification only involve persons as agents; in particular, voters are only expected
to communicate or otherwise interact with other persons, not devices such as checkers, scanners, copiers, software utilities, etc. 
Also, a manual recount is in principle possible anytime after election; we remark that the design does not assume the mapping of voters to their IDs will be archived for later recounts.

Various elements of the design may fail in two ways: such that will be detected at some stage of the election process, and others that will go undetected. The first, including errors in ID delivery, leaks in privacy thereof and subsequent invalidations of individual IDs, stalling of the ballot box, failure of the EC to read or accept semi-legible ballot papers as genuine, etc., will presumably incur complaints and may be detrimental to the validity of the outcome. We will however look at the design with especially the failures of the second type in mind.

In the proposal, it is easy to modify someone else's vote\footnote{This is an \emph{existential} statement:
there are persons who can easily modify some other persons' vote.}; it is difficult
 to do so in a meaningful way. 
The latter involves reading both parts and rewriting at least one.
Below, we discuss how voters put their trust in different roles in the design.

As for privacy, there are several separate pieces of private information:  
the voter's marked ballot paper; her ID; whether or not she voted by proxy, and the proxy's identity;
and possibly, whether she took part in the election.
The privacy of this information is maintained by distributing it.  
The mapping of voters to IDs is separate from the mapping of IDs to ballot papers; 
next to the separate delivery routes for the two parts of each ballot, 
the design tries to prevent it being evident whether a particular ballot paper was submitted by proxy.

\subsubsection{Background assumptions}

\smallskip\noindent
{\bf EC as a trusted authority.}
The electoral committee handle the delivered ballot papers and the list of valid ID's. 
The role of the EC in the proposed system mimics its role within widely used vote collection methods: 
as a body they derive their credibility from operating collectively 
under its putative heterogeneous nature with respect to polarities in the pool of voters.
The opportunity for error or fraud are reduced by this mode of operation;  
also, the delivered ballot papers are typically archived to enable a recount.
Physical manipulation (removing, replacing, or modifying) of the delivered ballot papers is difficult, plus, under the distributed nature of ballot papers proposed here, a purposeful modification is comparatively more difficult, given that the EC are not aware of the valid IDs in advance of the actual reading and counting of the ballots (step 4). 

\smallskip\noindent
{\bf RA as a trusted authority.}
The registering authority define---and hence know---both the list of valid IDs and
the bijection between voters and their IDs.  
Moreover, they can access the grid part of each voter's ballot paper;
in fact, each of the voters needs to submit her grid through the RA, no matter whether or not she visits the polling place.
So the RA keep one of the key roles in the election, one that is either new or significantly 
enhanced in the proposed system in comparison to a standard system,
and a correct and fair performance affects both voter privacy and correct voting intent expression.
As with the EC, the RA may operate as a body representing the pool of voters;
we do not delve into imaginable technical solutions within this role,  
 seeking rather to describe the interface between RA and the rest of the design.

Within an organization, the role of RA might be performed by the IT department.\footnote{This
was the setup in an earlier version of this proposal, \cite{Hanikova:BPV-techreport}.}
Under this scenario, it is the case that the persons in the role already act 
as a trusted authority there anyway, 
in matters of importance usually surpassing that of elections in that organization.
In particular, the IT could intercept any electronic communication to, from, or within the organization.

The communication between the RA and the voters is neither encrypted nor anonymous 
and is carried out in person or by usual communication means (email). 

\smallskip\noindent
{\bf Email is secure.} Email, perhaps alternated with other distant communication means, 
is essential to this design in that it caters for absentees' votes delivery, via the RA and the proxies. Email communication \emph{per se} is naturally not secure; let us look at how this affects the different parts of the design. Only the grid part of each ballot is uniformly delivered via the RA interface, and this delivery may or may not happen via email, depending on the setup and the preferences of individual voters. Interception of IDs is possible if delivered by email: 
whether on individual basis or \emph{en masse}, attempts at vote manipulation or alteration will be detected through invalidation of the affected IDs at vote count stage. An interception of IDs without attempts to manipulate the ballots may lead to undetected voter privacy infringements; leaking the IDs gives the receptors of this information a position comparable to that of the RA. 

It is harder, however, to meaningfully modify votes by intercepting and manipulating email: to alter someone's ballot paper in a meaningful fashion, 
it is necessary to be able to access both the grid and the grille part. The latter
is inserted in the ballot box in person by the voter or her proxy; in case it travels via email,
the exact route (from the voter to her proxy) is determined by the voter, and as such it is non-uniform across voters and in time, therefore imaginably difficult to predict and intercept. 

So the assumption of ``email security'', as made here, 
goes hand in hand with only some parts of the design relying on predictable email communication,
while other parts need not use email or may use it nonuniformly.

\smallskip\noindent
{\bf Proxy as a trusted party.}
Even if RA is a trusted authority and email is secure, it is perhaps hard to accept that 
Victoria ought to put her trust in one of her peers; 
not only Victoria must to be blessed with a clear-headed, trustworthy peer, 
but he needs to be able to come to the polling place.

Suppose that Victoria is lucky. Her proxy is helpful, smart and honest, 
so he will do his level best to record the grille part of her ballot faithfully
and in such a way that it will imply neither Victoria nor himself, and he will succeed in doing so. 
Moreover, he will reveal to no one that Victoria voted by proxy and that her proxy was himself, 
and he will not attempt to restore her ballot paper by somehow getting hold of the grid part. 
In this case, Victoria's vote is cast as intended, remains secret, and the fact that she voted by proxy remains hidden. 
Eventually, Victoria learns from the published results that her ID was included in the tally, although
she has no way of finding out, except for singular cases, whether her vote was cast as intended.

Suppose that Victoria is unlucky, and has placed her trust in a malevolent, forgetful, or silly proxy.
The proxy may fail to cast her vote, with or without intent. 
(This will transpire as soon as the results are announced, but 
cannot be rectified, and the effect is that Victoria will not have  participated in the election.) 
The proxy may further deanonymize Victoria's grille, e.g.,
printing her emailed grille including the header with Victoria's name in it, 
and inserting it in the ballot box as printed. 
(This will impinge on Victoria's privacy by revealing her vote to the EC 
but it will neither spoil her vote nor divulge her ID before the poll has closed.) 
Worse, the proxy may make public Victoria's ID, e.g., by leaving 
copies of Victoria's printed email lying loose. 
Under such scenario, not only the proxy but anyone learning the ID can create fake grilles for that ID 
and insert them in the ballot box. (This spoils Victoria's vote, as her ID will be invalidated.)
The proxy can swap Victoria's grille for another one. 
(This will go undetected by the EC if he preserves the checksum, 
and by anyone else bar singular cases.) 
Finally, the proxy may introduce errors in Victoria's grille.

In which manner can the proxy alter Victoria's grille? 
If he has no info about the grid part, he can only make random alterations. 
That does not help the proxy's cause (assuming he has one), because he does not know who he is in fact voting for, 
but it destroys Victoria's original intent. Neither Victoria nor the committee will detect the fraud, 
and everyone will have to live with the (altered) election result. 

With no trusted peer in sight, Victoria can improve her odds by choosing an election candidate as a proxy. 
The reasoning behind this is a putative correlation between trusting a person as a proxy, and an intended for-vote for that person.\footnote{This
perception was expressed in much stronger terms in \cite{Dodgson:Principles} in the context of parliamentary elections in the UK when dealing
with surplus votes, an issue irrelevant here. The author proposes that the candidate himself 
ought to be able to dispose of surplus votes for himself as he chooses; the reasoning goes as follows:
\emph{The Elector must understand that, in giving his vote to A, he gives it him as his absolute property, to use for himself, or to transfer to other Candidates, or to leave unused. If he cannot trust the man, for whom he votes, so far as to believe that he will use the vote for the best, how comes it that he can trust him so far as to wish to return him as Member?}}
The candidate might hesitate to tamper with a perceived for-vote for himself.
This might be seen as a voluntary privacy violation (Victoria hints to a candidate---truthfully or not---that she votes for 
him by asking him to act as her proxy), apparently not uncommon as things stand. 

The decision lies with Victoria between voting in person (if able to), 
voting by proxy, and abstaining from voting in that election. 
If she opts for a proxy, it is her choice entirely whom to appoint.

\subsubsection{Q\&A}
\label{sssection:qa}

\smallskip\noindent
{\bf Who knows what, once again?}
The RA define the valid IDs, and the pairing between IDs and voters.
Moreover, before the election day, the RA have assembled the grid parts of participating voters' ballots.
However, the RA cannot access the grille parts of voters' ballots, except possibly for their own ones.
Each person acting as a proxy for a voter learns the ID and the grille part of that voter. 
The EC have full access to the marked ballots under their respective IDs, 
but they do not know how IDs map to voters.   

Some information can be assumed to be generally known or at least potentially available. 
For example, it can be relatively easy to find out who actually visited the polling place, or on the other hand,
who was prevented from doing so. 
It is therefore important that there be no \emph{a priori} difference in appearance
between ballot papers submitted in person and those submitted by proxy: 
were that not the case, the voters who used a proxy might be implicated.

\smallskip\noindent
{\bf Why are some IDs invalidated?}
Some IDs may be invalidated for good reasons, usually indicating fraud. 
It is considered better to discard some votes that were subject to fraud than to discard the whole election.
Fraud can occur more easily with absent voters than to voters who come to polling place;
on the other hand, voters who are absent (or more generally, voters who use a proxy) can
contribute to minimizing the chance of fraud through their own choices. 
There is a fair expectation that absent voters will be able to cast a vote as intended.

\smallskip\noindent
{\bf Why can the ballots list more names than there are candidates?}
The possibility of expanding the list of candidate names is given to any voter in order to hide from 
her proxy how many for-votes she has given to real candidate names. 
Twice the number of candidates ought to suffice to hide
any choice among real candidates. 

\smallskip\noindent
{\bf Is there a limit on ballot box access?}
Each person (a voter or a proxy) is expected to insert 
one grille for herself and up to $n-1$ grilles for her peers ($n$ being the number of voters),
if she happens to be acting as proxy for them. 

However, the design does not expect the EC to keep track of ballot box access.
The lack thereof implies that, at any given time, 
 the EC may only have a vague idea how many papers each person has inserted;
and even if they do have an idea, there is no easy way to prove that they are right.
Thus the EC effectively cannot block access to the box.
 
Disruptive behaviour, whether a prank or a test of the system,  
is of course a trial to the EC,
both on social and on practical level.
Going through heaps of papers to find the few ones with a valid ID will inconvenience the EC,
although it will not invalidate the election result.
Should disruptions be anticipated, the committee  might impose (ahead of the election) some access control,
perhaps limiting the pieces of paper up to $n$, for example; 
this would necessitate ballot box access control in the manner of the standing system.

\smallskip\noindent
{\bf Is it courteous to refuse to act as a proxy?}
Acting as someone else's proxy involves both some effort and a good deal of trust that goes both ways. 
The effort rests in a careful transmission of the ballot information, and perhaps some care in not mentioning it.
More importantly, if asked to act as a proxy for Victoria, one needs to be reasonably sure
that the information that Victoria submits through oneself is her genuine vote, not random noise generated by her.
Yet another possible concern is that one might be asked to act as a proxy while having no intention of
either voting at all, or voting in person (so one cannot act as a proxy for another voter).
Thus acting as as someone's proxy is by no means a matter of course, refusals may be curt, and will bear no discussion.

\smallskip\noindent
{\bf Can one prove how one voted?}
One can prove how one voted to the EC, by submitting to them one's ID 
before \emph{all} IDs are made public. 
In the standard system, a rough analogy would be marking one's ballot paper in some special way and confronting 
the EC with the information. 

\smallskip\noindent
{\bf Can one complain?}
There is no protocol to credibly file a complaint.
In particular, for Victoria who voted by proxy, there is no chance to trace her vote and 
no occasion to rectify her spoiled vote,
should it occur.

\smallskip\noindent
{\bf How large should the ID numbers be?}
The default setting assumes no box access control: 
everyone who can access the box can insert whatever they please. 
The IDs must be large enough to sustain an attempt to 
guess, and possibly invalidate, some of them simply by stuffing ballots under different numbers.

Shortly put, the IDs must be proportional in size to the physics of the election.
As stated already, the physics of this proposal is the manual one:
a voter inserts sheets of paper through a slot in the lid of a box.
The magnitude of the IDs ought to make the possibility of invalidation negligible
under a scenario of one or more persons bringing in fake ballot papers 
at the limits of their physical ability.

Whatever range the IDs are actually chosen from, 
there appears to be no need for the RA to announce the range to anyone.

\section{Modifications}
\label{section:modif}

The modifications suggested in this section ought to be viewed as independent of each other,
all of them modifying the proposed system as presented. Some of them can be combined.

\subsection{Different electoral systems}

This proposal is suitable for several electoral systems, such as 
first past the post, 
plurality at large (including the condition that no voter can give more votes than there are seats),
or ranked voting. 

Under some of these systems (such as first past the post), the number of for-votes given by a single voter 
is fixed; it is therefore not necessary to hide it from a proxy, and each ballot will list only the real candidate names.

\subsection{Two proxies}

This variant lessens the role of RA somewhat. Each voter still receives her ID from the RA.

Victoria, intending to participate in an election, creates a two-part ballot as above. 
Provided she attends in person, she inserts both parts in the ballot box on election day;
it is no longer necessary that any part of the ballot of a voter pass through the  RA. 
If not voting in person, she uses two different proxies who are not informed of each other's identity, but 
both agree to be present at polling place. (One of the two can still be the RA.)
As a gain, voters who attend in person need not delegate any part of their ballot to anyone.

A minor issue is that a ballot paper that the voter physically created herself may bear unwanted characteristics,
indication not only that the grid-grille pair form a ballot (which is fine, and is indicated by their ID) but also that
neither has been submitted by proxy (who necessarily alter the physical nature of the ballot).
In fact, voters might prepare their ballots literally by writing two columns on a piece of paper and separating them later;
the problem is leaving it at that.
If most voters who attend in person act this way, the absent voters
are implicated. 
Perhaps additional caution is needed so that the physical nature of ballots does not indicate 
whether the voter had a proxy.

\subsection{Default ballot papers}

Some voters might prefer not having to create their ballot papers from scratch.
This can be achieved in a number of ways that do not alter the design 
while increasing voter comfort. For example:

\begin{itemize}
\item \emph{Default grid.} With the ID, the RA deliver to a voter a list of names, of fixed length $2n$, 
containing all of $n$ candidate names and some other names, in a random order.
(The order, and the position of the real candidate names, is specific and private to that voter.) All the voter needs to do is to 
assign her preferences by preparing the grille part. The voter will either confirm this default grid, or discard it and return one of her own to the RA.
\item \emph{Default grille.} In this variant, the submission routes of the grid and the grille are swapped. 
With the ID, the RA deliver to a voter a $0/1$ vector of length $2n$, with exactly $n$ $1$'s, in random order private to the voter.
The voter expresses her preferences by assigning candidate and other names to this default grille. 
The voter can discard the default
and substitute her own grille to the RA.
\end{itemize}

\subsection{A numerical encoding}

To abstract from the candidate order, and the associated risks of a mistake in copying the vector,
one can consider coding candidate names by numbers. The grid part, which the voter prepares first, consists of real or imagined  candidate names, 
and for each name, a unique random natural number. (A default grid of this nature can also be supplied by the RA.)
The voter returns the grid to the RA. The grille part, which is submitted in person or by proxy to the EC, contains the numbers of the voted-for candidates.


\subsection{A security embellishment}

Victoria cannot prevent her grille being altered by her proxy.\footnote{It would not 
help much, for example, if by design Victoria could use more proxies (perhaps not aware of each other) 
to submit her one grille. 
The scheme would also necessitate different set of rules for the EC
to accept delivered grilles; namely, they would have to accept multiple papers with the same ID provided they all specified
exactly the same grille. However, if the multiple proxies delivered 
differing grilles under Victoria's ID, then the committee would have to invalidate that ID, 
so such a modification would in fact introduce more vulnerability by relying on more than one person's honesty.}
And not only cannot she prevent fraud; she need not notice it if it occurs. 

To give Victoria a better chance of confident absentee vote, a small security embellishment can be added.
Along with her grid and checksum that she returns to the RA before election day, 
Victoria can submit additional information about the grille to the RA.
This will be passed on to the EC, 
who will thus be able to evaluate whether or not Victoria's original grille has been delivered.
The nature of such information is not specified here. 
It needs to be simple enough for the  committee to evaluate without undue effort.
Moreover, again so as not to implicate the absent voters, \emph{all} voters ought at least consider including
some additional information even if they vote in person.

This does not give Victoria a better chance of preventing alterations to her ballot, 
it only increases the chance that any changes will be detected by the EC,
whereupon her ID will be invalidated.
The difference, therefore, is that if her grille had been tampered with, Victoria will not have participated
in the election, rather than having cast a random vote.

\subsection{Swapping IDs}

The fact that each vote is cast under an ID that maps uniquely to a voter can be perceived as a major hindrance.
While this is only a problem in case the IDs are intentionally leaked, it may cause discomfort, 
and various  measures devised to alleviate it are imaginable. 

Voters can however, on an individual level, oppose
the \emph{big brother} effect of the IDs by swapping them privately. A pair of voters agree
to reveal their IDs to each other;  they swap their grids (since these contain no meaningful info, they can
simply return each other's grid to the RA instead of their own) and subsequently cast their vote by inserting
their own grille marked with their party's ID.

This is a daredevil game to a large extent, not changing the overall properties of the design
but subverting it locally.  

\subsection{Transitive proxy}

By a \emph{transitive proxy} we mean a modification of the design where the person appointed as a proxy 
can delegate this responsibility further, i.e., transmit the grille and ID, originally entrusted to himself by Victoria,
to another proxy (possibly along with his own vote, if, for example, it turns out that he is not able to attend 
the election in person after all).


Despite all its imaginable pitfalls, this scheme seems more robust than the default one.
It is also, by its nature, not detectable, and may occur even if not explicitly allowed.
It ought to be made clear to voters prior to an election whether 
transitive proxy is a part of the official policy. 

\subsection{Publishing the ballots}

After an election (step 4), optionally the EC may publish a ``purged'' version of valid delivered ballot papers.
A purged ballot paper is the list of candidate names that have been voted for from the original ballot paper,
without its ID and without any additional names that may have been included by the voter.\footnote{There is no doubt
that this optional move---which ought in any case be subject to approval of all voters before the commencement of the 
election---is a mixed blessing. First and foremost, the purged ballots are only informative if the candidate list is long enough: what is being published is the combinations of for-votes. 
While it might be taken as a supportive measure for fraud detection, and it might alert
the pool of voters to clubbing (by which we mean an unusual amount of identically marked ballot papers),   
one might object that publishing the combinations is already an infringement on privacy (cf.~the main objection
to the ThreeBallot system, \cite{Rivest:ThreeBallot}), in addition to burdening the EC with extra work.}

\section{Concluding remarks}

The blinded mechanism of proxy voting, offering privacy to the principal while enabling voting in absence,
 is the main import of this proposal. It comes at a cost: each vote is cast under an identification
number, and the assignation of numbers to voters is a sensitive information indeed. 
In addition, the election process is somewhat more laborious compared to the standard vote collection.
A minimal overhead is the involvement of the RA and the
assembling of grid-grille pairs by the EC. 
 
This is perhaps an instance of a more general phenomenon of the extra effort incurred 
by a majority in a group (voters who can vote in person) accommodating the needs of a minority
(absent voters). 
The tradeoff between simplicity of the solution and its universality is expectable.
It needs to be deliberated carefully whether voters are \emph{likely} to vote in absence under the
proposed design; that is, whether the extra effort is worthwhile.

We remark that, although not implicit in the standard method, it is far from exceptional that 
a voting mechanism imposes and maintains a link between a voter
and her ballot paper (or its number);
see the discussion in \cite{Jones:ProblemsE2EVoting}.

Grille encryption is notorious as a child's toy; this does not disqualify
it for the use proposed here, i.e., for blinding a proxy to the transmitted information.
An obvious advantage is that it is very easy to use.
The idea of permuting the names on a candidate list in order to hide voting intent
 is quite intuitive too: 
Pr{\^e}t \`a Voter (\cite{Ryan:PretAVoter}) uses such permutations, although it does not seem to offer a possibility of remote voting.


The proposed system is not auditable, and as such, it is suitable 
for communities where implicit trust in electoral authorities is  high.
It maintains the privacy of each voter, and contributes to it by voters being able to make 
choices for themselves within the system.

\section{Acknowledgements}

The work was partly supported by the long-term strategic development financing of the Institute of Computer Science, Czech Academy of Sciences (RVO:67985807). The author is indebted to her colleagues from the Department of theoretical computer science of said Institute for comments on earlier versions of the manuscript.

\bibliographystyle{plain}


\end{document}